\documentstyle[12pt]{article}
\def\setb@se#1{\baselineskip=#1 \normalbaselineskip=#1}
\setb@se{14pt}
\newcommand{\be}{\begin{equation}}
\newcommand{\ee}{\end{equation}}
\newcommand{\g}{{\bf g}}

\begin{document}

\input epsf

\begin{titlepage}
\begin{flushright}
ZU-TH 8/97

April 3, 1997

hep-th/9704026
\end{flushright}
 
\vfill

\begin{center}
{\huge  Slowly Rotating Non-Abelian 

\vspace{2 mm}

Black Holes}

\vfill

{\bf Mikhail S. Volkov\ and \ Norbert Straumann} 

\vfill

{\em Institute of Theoretical Physics, University of Z\"urich,}\\
{\em Winterthurerstrasse 190, CH--8057 Z\"urich, Switzerland}\\

\vfill

{\Large\bf Abstract}

\vfill

\end{center}

\noindent
It is shown that the well-known non-Abelian static SU(2) black hole
solutions have rotating generalizations,
provided that the hypothesis of linearization stability is accepted. 
Surprisingly, this rotating branch has an asymptotically 
Abelian gauge field with an electric charge that cannot vanish,
although the non-rotating limit is uncharged. 
We argue that this may be related to our 
second finding, namely that there are no globally regular slowly
rotating excitations of the particle-like Bartnik-McKinnon
solutions.

\vfill 

\end{titlepage}

\noindent
{\bf Introduction.--}
Ever since the discovery of a discrete family of particle-like solutions
of the Einstein-Yang-Mills (EYM) equations by Bartnik and McKinnon
\cite{1}, as well as of their black hole analogues by several workers
\cite{2}, \cite{2a}, it has been a challenge to find rotating generalizations
of these remarkable objects. This is an exceedingly difficult task. 
It is not even clear how to parametrize the metric, because 
the coupled field equations do {\sl not} imply the Frobenius
integrability conditions for the Killing fields \cite{3}. 
The usual Papapetrou ansatz \cite{4} may therefore be too narrow. 
Even if this ansatz is adopted, the resulting system of partial
differential equations becomes considerably more involved than in the 
Abelian case. 

In view of this situation it is reasonable to pursue in a first step a 
more modest goal, based on the assumption of linearization stability
\cite{stab}. 
Suppose that there is (at least) a one-parameter family of stationary
black hole or regular solutions of the EYM equations, 
approaching  the static solutions 
mentioned above for angular momentum $J=0$, 
then the tangent to this family at $J=0$ satisfies
the linearized EYM equations. Conversely, it is reasonable to expect that 
for a well-behaved solution of the linearized equations around the 
static configurations there exists an exact one-parameter family 
of stationary  solutions. 

In this paper we show that, under the above assumption of 
linearization stability, there are slowly rotating non-Abelian black holes
branching off from the (numerically known) static ones. 
To our surprise these black holes are ``charged up", in that they possess
an asymptotically Abelian gauge field with a non-vanishing
electric charge. A second result, which may also be surprising to
many workers in the field, concerns slowly rotating globally regular
solutions: It turns out that there are no acceptable rotational modes in this case;
all nontrivial ones develop singularities at the origin. 
At the end of the paper we shall give
a heuristic interpretation of how this may be related to the charging up
of the black holes. 
We can, of course, not exclude that there is a disconnected rapidly
rotating branch above some non-vanishing angular momentum. 

Recently, there have been a number of investigations \cite{5,6} on {\sl static}
axially symmetric solutions for matter models with YM fields which do not
exhibit spherical symmetry, but, to our knowledge, we present here the 
first results on rotational deformations.

\noindent
{\bf Perturbation equations.--}
The EYM action for the SU(2) gauge group reads
in standard notation
\be                                                       \label{1}
S=\int\left(-\frac{1}{4}R+\frac{1}{2}\, {\rm tr}\, F_{\mu\nu}F^{\mu\nu}
\right)\sqrt{-\g}\, d^4 x,
\ee
with $F_{\mu\nu}=\partial_{\mu}A_{\nu}-\partial_{\nu}A_{\mu}+
[A_{\mu},A_{\nu}]$. 
In the decomposition 
$A_{\mu}=A_{\mu}^{a}{\rm T}_{a}$ of the gauge potential 
we choose the gauge group generators ${\rm T}_{a}=\tau^{a}/2i$
with $\tau^{a}$ being the Pauli matrices.

An intensively studied 
family of non-Abelian black hole solutions \cite{2}, \cite{2a}
 is described by 
static, spherically symmetric metrics parametrized as
\be                                                          \label{2}
ds^2=\left(1-\frac{2m(r)}{r}\right)\sigma(r)^2 dt^2-
\frac{dr^2}{1-2m(r)/r} -r^2(d\theta^2+\sin^2\theta\, d\varphi^2),
\ee
and purely magnetic gauge fields of the form
\be                                                          \label{2:0}
A=w(r)(-{\rm T}_{2}\, d\theta+{\rm T}_{1}\sin\theta\, d\varphi)+
{\rm T}_{3}\cos\theta\, d\varphi. 
\ee 
The functions $w(r)$, $m(r)$, and $\sigma(r)$ 
are subject to the EYM field equations.
The solutions
are characterized by the event horizon radius $r_h$ and an integer $n$
counting the number of nodes of $w$.
We shall need the asymptotic behavior of the solutions,
\be                                                     \label{2:2}
w=w(r_h)+O(x), \ \ m=\frac{r_h}{2}+O(x);\ \ \ \ \ \ 
w=-1+\frac{a}{r}+O\left(\frac{1}{r^2}\right),\ \ 
m=M+O\left(\frac{1}{r^3}\right),
\ee
at the horizon and infinity, respectively; here $x=r-r_h$, and
$w(r_h)$, $a$, and $M$ are fixed by the solution
parameters $r_h$ and $n$; the behavior of $\sigma$ is 
determined by that for $w$.

Consider perturbations of a given background equilibrium solution
$(\g_{\mu\nu},A_{\nu})$,
\be                                                        \label{3}
\g_{\mu\nu}\rightarrow\g_{\mu\nu}+\delta\g_{\mu\nu},\ \ \ \
A_{\nu}\rightarrow A_{\nu}+\delta A_{\nu}.
\ee
The perturbation equations are obtained by  linearization
of the field equations:
\be                                                         \label{4}
\delta R_{\mu\nu} = 2\, \delta T_{\mu\nu},\ \ \  \ \ \ \ 
\delta\left( D_{\mu}F^{\mu\nu}\right) =0,
\ee
where $D_{\nu}\equiv \nabla_{\nu}+[A_{\nu},\ \cdot \ ]$.
With the notations $h_{\mu\nu}\equiv\delta\g_{\mu\nu}$ and
$\psi_{\mu}\equiv\delta A_{\mu}$ the perturbation equations
take the form
$$                                                    
-\nabla_{\sigma}\nabla^{\sigma}h_{\mu\nu}-
2\, R_{\mu\alpha\nu\beta}h^{\alpha\beta}+
R^{\sigma}_{\mu}h_{\sigma\nu}+
R^{\sigma}_{\nu}h_{\sigma\mu}=
4\, \delta T_{\mu\nu},
$$
\be                                                     \label{5}
-D_{\sigma}D^{\sigma}\psi_{\nu}+R^{\sigma}_{\nu}\psi_{\sigma}
-2[F_{\nu\sigma},\psi^{\sigma}]+
h^{\alpha\beta}D_{\alpha}F_{\beta\nu}+
F^{\alpha\beta}\nabla_{\alpha}h_{\beta\nu}=0,
\ee
provided that the following gauge conditions are imposed: 
\be                                                      \label{6}
\nabla_{\sigma}h^{\sigma}_{\nu}=h^{\sigma}_{\sigma}=
D_{\sigma}\psi^{\sigma}=0.
\ee
The quantity $\delta T_{\mu\nu}$ in Eqs. (\ref{5}) is obtained by varying 
the energy-momentum tensor 
\be                                                      \label{7}
T_{\mu\nu}=\frac{1}{2}{\rm tr}\left( 
F_{\mu\alpha}F_{\nu\beta}\, \g^{\alpha\beta}
-\frac{1}{4}\g_{\mu\nu}
F_{\alpha\beta}F_{\rho\sigma}\g^{\alpha\rho}\g^{\beta\sigma}\right)
\ee
with respect to the metric and the gauge field. 

\noindent
{\bf Rotational perturbations.--}
To identify the most general rotational degrees of freedom, 
we determine those amplitudes in the partial wave decomposition
which can give a non-vanishing contribution to the ADM flux 
integral for the total angular momentum
\be                                                        \label{7a}
J^i=\frac{1}{32\pi}\oint_{S^2}\varepsilon_{ink}\, 
(x^k\, \partial_j h^{0n}+\delta^n_j\, h^{0k})\, d^2 S^j. 
\ee
We do this by expanding 
$h_{\mu\nu}$ and $\psi_{\mu}$ with respect to an
appropriately chosen complex null tetrad.  (This approach
has proved to be extremely efficient
for solving wave equations \cite{chandra}.)
The complete separation of the angular variables
in Eqs. (\ref{5}) is achieved, provided that each of the tetrad
projections (for $\psi_{\mu}$ we project in addition onto
${\rm T}_{\pm}={\rm T}_1\pm i{\rm T}_2$, and T$_3$)
is chosen as the product of a radial amplitude,
depending only on $t$ and $r$,
and a spin-weighted spherical harmonic
$_sY_{lm}(\theta,\phi)$ \cite{spin}. Here the spin weight
is $s=0,\pm 1,\pm 2$,  depending on the 
projection under consideration. 
 Passing back to the coordinate basis, we obtain the full
mode decomposition for perturbations
with respect to the angular momentum quantum numbers. 
(We do not present here the explicit expressions in view of their
complexity.)

Since the background solutions are spherically symmetric, group
theoretical arguments imply that modes with different pairs 
$(l,m)$ decouple, and for a given $l$ (an irreducible representation
of the rotation group) there is a $(2l+1)$-fold degeneracy
labeled by $m$. The angular order $m$ does not occur in the 
equations for the radial amplitudes, and we can thus choose $m=0$. 
Next, passing to cartesian
coordinates $x^i$, we compute the ADM angular momentum 
by integrating   over a two-sphere 
at finite radius $r$ in Eq. (\ref{7a})
and then taking the limit $r\rightarrow\infty$. 
The angular dependence of the integrand implies then that
the integral vanishes for any $r$  unless $l=1$, and that
only the $h_{0\varphi}$ perturbation
component can give a non-vanishing contribution. 
Choosing $l=1$ and suppressing the
time-dependence, the tensor modes with the spin weight 
$s=\pm 2$ vanish and Eqs. (\ref{5}) reduce to a system of 
18 coupled equations. 

Now,  the transformation behavior of the angular momentum
under space and time reflections (P,T) implies that only those
perturbation amplitudes are relevant which are even under P
and odd under T. For $l=1$ these appear only in 
$h_{0\varphi}$ and in two isotopic component of $\psi_{0}$. 
They decouple from the remaining modes because the 
background solutions are P and T symmetric. 
This analysis leads finally
to the following {\sl most general 
ansatz} (up to global coordinate rotations) 
for the stationary rotational modes:
\be                                                    \label{8}
h=2S(r)\sin^2\theta\,  dt\, d\varphi,\  \ \
\psi=\left({\rm T}_{1}\frac{\chi(r)}{r}\sin\theta+
{\rm T}_{3}\frac{\eta(r)}{r}\cos\theta\right) dt\, . 
\ee
(The invariance of $\psi$ and the background 
gauge field (\ref{2:0}) under P becomes manifest
after a suitable gauge transformation). 
Conditions (\ref{6}) for the ansatz are fulfilled identically
and the perturbation equations (\ref{5}) reduce to the following coupled
system for the radial amplitudes $S$, $\chi$, and $\eta$ in (\ref{8}): 
$$
-r^2 N\sigma\left(\frac{S'}{\sigma}\right)'+
\left(2N+4\frac{(w^2-1)^2}{r^2}\right)S
+4Nr^2 w' \left(\frac{\chi}{r}\right)'+
\frac{4(w^2-1)}{r}\left(w\, \chi-\eta\right)=0,
$$
$$
-r^2 N\sigma\left(\frac{\chi'}{\sigma}\right)'+
\left(1+w^2-2w'^2 N\right)\chi
- 2w\,  \eta- rN \left(w'S\right)'
+\left(2N w'^3 +\frac{w(w^2-1)}{r}\right)S=0,
$$
\be                                                 \label{9}
-r^2 N\sigma\left(\frac{\eta'}{\sigma}\right)'+
2\left(1+w^2- w'^2 N\right)\eta
- 4w\, \chi+\frac{2(1-w^2)}{r}\, S=0.
\ee
Here $w$, $m$, and $\sigma$ refer to the background solutions,
$N\equiv 1-2m/r$. For a solution of these equations, the ADM
angular momentum is
\be                                                              \label{9a}
J^i=\delta^i_z\, 
\lim_{r\rightarrow\infty}\frac{1}{6}\, r^4\left(\frac{S}{r^2}\right)'.
\ee

\noindent
{\bf Qualitative considerations.--}
Before turning to the numerical analysis of the radial
equations, we make some qualitative remarks.  Consider first a 
Kerr-Newman black hole with  mass $M$,
angular momentum $J$ and electric charge $Q$. Let  ${\cal A}$ denote
the electromagnetic potential of the solution. 
Next, consider the embedding of this Abelian solution into the SU(2) gauge theory, 
such that the gauge field potential is $A={\cal A}\, {\rm T}_3$. 
Suppose that $|J|$ and $|Q|$ are small and linearize the solution 
with  respect to $J$ and $Q$. The result can be viewed as a
Schwarzschild black hole with two linear hairs. The axial gravitational
perturbation can be represented in the form of Eq. (\ref{8}) 
with 
\be                                                \label{10}
S(r)=-\frac{2JM}{r}, 
\ee
while the Yang-Mills hair is described by $A={\rm T}_3 (Q/r) dt$. After a 
gauge transformation with the SU(2)-valued function 
$U=\exp((\pi-\theta){\rm T}_2)\exp(\varphi{\rm T}_3)$
this gauge field becomes 
\be                                                 \label{11}
A = 
\frac{Q}{r}\, U{\rm T}_3U^{-1}+UdU^{-1}=\psi + A_{\rm pure}, 
\ee
where $A_{\rm pure}$ is a pure gauge potential of the form 
 (\ref{2:0}) with $w(r)=-1$, and the perturbation $\psi$ is given by
\be                                                      \label{12}
\psi=\frac{Q}{r}\left({\rm T}_{1}\sin\theta-
{\rm T}_{3}\cos\theta\right) dt. 
\ee

Now let us return to the perturbation equations (\ref{9}) and concentrate on 
their solutions in the asymptotic region $r\gg r_h$. In this limit, the 
background geometry is approximately 
Schwarzschild and $w(r)\approx -1$, i.e.,  the background gauge field 
is almost pure gauge. Eqs. (\ref{9}) then split into the two independent
groups -- one equation for the metric perturbation $S$, and two 
coupled equations for the Yang-Mills amplitudes $\chi$ and $\eta$. 

The acceptable solution of the first equation is  given by Eq. (\ref{10}). 
The equations for $\chi$ and $\eta$ admit the solution $\chi(r)=-\eta(r)=Q$
with constant $Q$, which, together with the background pure gauge field, 
exactly corresponds to the gauge field specified by Eqs. (\ref{11}), (\ref{12}). 
We therefore conclude that the solution of the perturbation equations
in the asymptotic region is close to the linearized Kerr-Newman
solution with charge $Q$ and angular momentum $J$. 

So far the parameters $J$ and $Q$ are independent. 
However,  Eqs. (\ref{9})   split into the 
two independent groups only asymptotically. It is to be  expected
that the regularity of the 
solution in the entire domain $r\geq r_h$
will imply relations for 
the parameters $J$ and $Q$ describing the asymptotics. 
We shall indeed show below that 
the charge $Q$ of the black hole is 
 uniquely fixed by the value 
of its angular momentum $J$.

\noindent
{\bf Numerical solutions.--} We now describe briefly the numerical 
analysis of the full problem. 
Guided by the Abelian example,  we are looking
for global solutions of Eqs. (\ref{9}) which are everywhere regular. 
Note that they do not have to be normalizable, as the Kerr-Newman
example already shows. 
Eqs. (\ref{9}) have regular singular points at the horizon
and at infinity. At the horizon, the 
formal power-series solution is found to be (using Eq. (\ref{2:2})):
\be                                                         \label{13}
S=c_{0}+c_{1}\, x+O(x^2),\ \
\chi=-\frac{c_{0}\, w(r_h)}{r_h}+c_{2}\, x+O(x^2),\ \
\eta=-\frac{c_{0}}{r_h}+c_{3}\, x+O(x^2),
\ee
where  $c_{0}$, $c_{1}$,   $c_{2}$, and $c_{3}$
are four independent integration constants.
The solution at infinity contains three independent parameters,
$J$, $Q$, and $c_{4}$: 
$$
S=-\frac{2JM}{r}+\frac{aQ}{r^2}+O\left(\frac{1}{r^3}\right),\ \
\chi=Q\left(1+\frac{a^2}{15}\frac{\ln r}{r^2}\right)+
\frac{c_{4}}{r^2}+O\left(\frac{\ln r}{r^3}\right),
$$
\be                                                       \label{14}
\eta=Q\left(-1+\frac{2a^2}{15}\frac{(\ln r-5)}{r^2}  \right)
+\frac{2c_4}{r^2} +O\left(\frac{\ln r}{r^3}\right). 
\ee
Clearly,  $J$ and $Q$ are, respectively, the angular 
momentum and the charge of the black hole. 

Now, we extend these asymptotic solutions from both sides 
to the intermediate
region and impose the matching conditions for the
functions $S$, $\chi$, and $\eta$, and their derivatives.
This gives six linear algebraic equations for the
seven free parameters in (\ref{13}), (\ref{14}), and hence
the matching can generically be fulfilled. Thus
there exists globally a solution whose asymptotic behavior
is specified by Eqs. (\ref{13}) and (\ref{14}).
For a given $J$  the values of the remaining
six coefficients will be proportional to $J$. 
In particular, we have a relation of the type
\be                                                         \label{15}
Q = \Omega_n (r_h)\,  J,
\ee
where $\Omega_n (r_h)$ is a function determined by the 
background non-Abelian black hole solution. 

\begin{figure}
\epsfxsize=8cm
\centerline{\epsffile{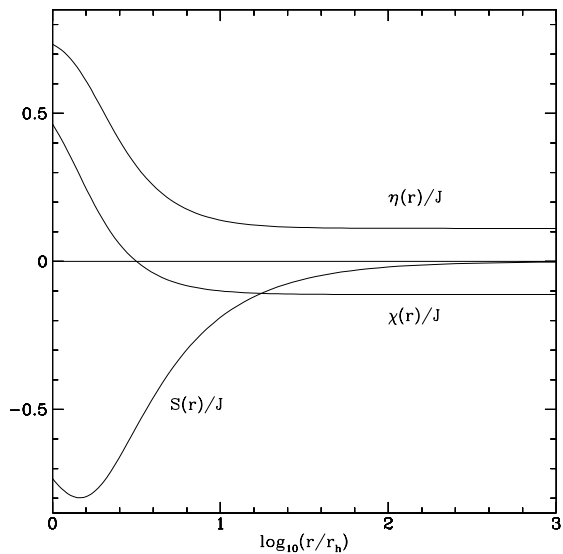}}
\caption{
The radial behavior of the rotational
excitations 
for the $n=1$, $r_h=1$ non-Abelian
black hole solution. }
\label{Fig.1}
\end{figure}

The numerical integration of Eqs. (\ref{9}) (see Fig.1) 
with the boundary conditions specified by Eqs.(\ref{13}), (\ref{14})
reveals that in  general none of the seven coefficients
in (\ref{13}), (\ref{14}) vanishes. We therefore
conclude that, under the assumption of linearization stability, 
each non-Abelian black hole solution admits
 stationary rotational generalizations with
electric charge given by Eq. (\ref{15}). 
The behavior of the function $\Omega_n (r_h)$
for the lowest $n$'s and for $0<r_h<\infty$
is shown in Fig. 2; numerical values are given in Table 1. 

\begin{figure}
\epsfxsize=8cm
\centerline{\epsffile{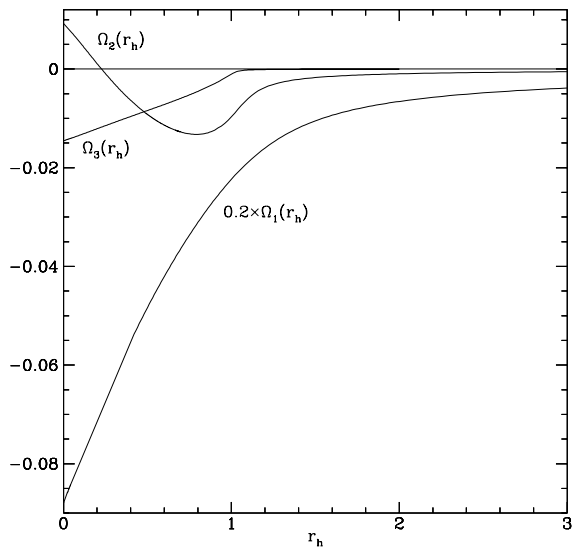}}
\caption{The charge-to-angular-momentum ratio
$Q/J$ versus the event horizon radius 
for the three lowest branches of rotating black holes
solutions.
}
\label{Fig.2}
\end{figure}

\begin{center}
Table 1. Numerical data for $\Omega_n(r_h)$.
\vglue 0.4cm
\begin{tabular}{|c|c|c|c|c|c|c} \hline
$n$ 
& $r_h=0$ 
& $r_h=0.5$ 
&$r_h=1$ 
& $r_h=5$    
&$1/\alpha\equiv r_h\gg 1$            \\ \hline
1
&  $-0.4396$  
& $-0.2416$  
& $-0.1117$ 
& $-0.0108$ 
& $-0.0052\, \alpha+O(\alpha^2)$\\
2
&   $+0.0092$   
&   $-0.0091$
&   $-0.0096$
&   $-0.0003$
&  $-0.0014\, \alpha+O(\alpha^2)$       \\
3
&   $-0.0145$
&   $-0.0084$
&  $-0.0010$
&  $-8\cdot 10^{-6}$
&  $-3\cdot 10^{-5}\, \alpha+O(\alpha^2)$     \\
\hline
\end{tabular}
\end{center}

\noindent
{\bf Discussion.--}
To obtain the asymptotic behavior of $\Omega_{n}(r_h)$ for 
$r_h\gg 1$ we make use of the fact that, in this limit, 
the spacetime geometry of the non-Abelian black holes
is Schwarzschild, up to  corrections
of  order $1/r_h^2$. 
Then,  introducing the new radial coordinate $\xi=\alpha\, r$
with $\alpha\equiv 1/r_h$, in terms of which the  metric functions 
are
$N=1-1/\xi+O(\alpha^2)$, $\sigma=1+O(\alpha^2)$, 
we expand Eqs. (\ref{9})
with respect to
$\alpha$. The solution in zeroth order of the expansion is
$S^{(0)}=-J/\xi$, and $\eta^{(0)}=\chi^{(0)}=0$. 
In first order, we obtain two 
linear equations for $\eta^{(1)}$ and $\chi^{(1)}$ 
with the source terms constructed from $S^{(0)}$. 
These equations, as Eqs. (\ref{9}), 
admit  solutions which become constant 
at infinity, thus giving rise to an electric charge of order
$\alpha$.

In the opposite limit, $r_h\rightarrow 0$, the functions
$\Omega_n(r_h)$ assume finite values. 
The non-Abelian black hole solutions 
converge 
for $r_h\rightarrow 0$ 
in the region $r> 0$ pointwise to the 
regular Bartnik-McKinnon solutions \cite{1}, 
which suggests that the same holds
for their perturbations. In other words, one can expect that
the particle-like solutions admit rotational excitations as well.
We have to investigate, however, what happens at the origin. 
There, the behavior of the unperturbed solutions is
$w=\pm (1-\beta r^2+O(r^4))$, 
$m=2\beta^2 r^3+O(r^5)$, $\beta$ being a parameter. 
Eqs. (\ref{9}) then give for the perturbations
$$
S=-4\,  a_0\, \beta \left(r-\frac{\beta}{5}\left(1+4\beta^2\right)r^3\right)
+a_2\, r^2+O(r^4),
$$
$$
\chi=a_0\, \left(1-10\beta^2 r^2\right)+a_1\, r+
\left(a_2\, \beta-\frac{a_3}{2}\right)r^3+O(r^4),
$$
\be                                                      \label{16}
\eta=a_0\, \left(1-8\beta^2\, r^2\right)+a_1\, r+a_3\, r^3+O(r^4),
\ee
where $a_0$, $a_1$, $a_2$, and $a_3$ are four integration
constants. Notice that the number of free
parameters, together with those in the asymptotic solution
(\ref{14}), is again large enough for matching in the
intermediate region. The numerical integration of Eqs. (\ref{9})
for the regular backgrounds with the boundary conditions given by
(\ref{14}) and (\ref{16}) shows that none of the seven
parameters in (\ref{14}), (\ref{16}) vanishes, and the 
solutions in the region $r>0$  are indeed very close
to the black hole rotational modes
obtained in the limit $r_h\rightarrow 0$. 
In particular, the ratio $Q/J$ for the 
particle-like solutions is given by $\Omega_n(0)$. 
However, although the functions $S$, $\chi$ and $\eta$ are 
regular, the corresponding perturbations (\ref{8}) 
nevertheless become singular
at the origin, since
the field ansatz (\ref{8}) involves the factor $1/r$, while 
the coefficient $a_0$ in Eq. (\ref{16}) does not
vanish. We therefore conclude that the regular solutions
do not admit rotational excitations.

\begin{figure}
\epsfxsize=8cm
\centerline{\epsffile{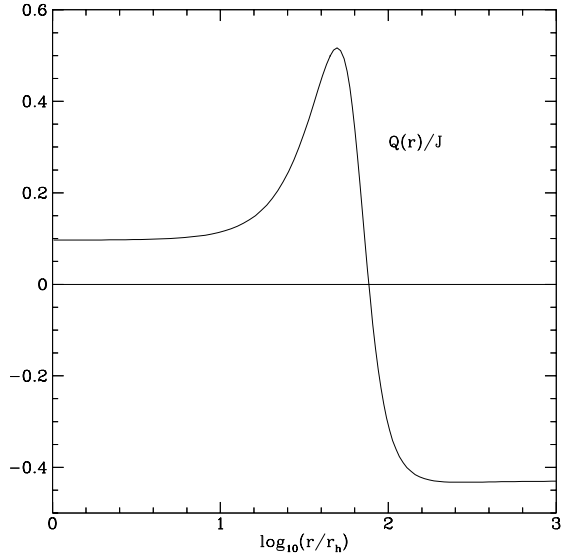}}
\caption{The charge function Q$(r)$ 
given by Eq. (21) for the $n=1$, 
$r_h=0.22$ black hole
}
\label{Fig.3}
\end{figure}

The appearance of a nonvanishing electric charge
induced by the rotation is a somewhat surprising feature of the
solutions under consideration -- the static non-Abelian black
holes being neutral. It is natural to wonder where this
charge originates from and where it resides. 
These questions cannot be answered in a gauge invariant manner,
but it may be instructive to compute the flux 
\be                                               \label{17}
I=\oint_{S^2} \ast\delta F
\ee
through a two-sphere of finite radius
in the distinguished gauge where the field $\delta F$ becomes 
asymptotically Abelian.  This gives uniquely 
\be                                               \label{18}
I=-{\rm T}_3 \oint_{S^2}
\ast\left\{\left(
-\left(\frac{\chi}{r}\right) '\sin^2\theta\,+
\left( \frac{\eta}{r}\right) '\cos^2\theta\,
\right) dt\wedge dr\right\},
\ee
which motivates us to introduce the charge function 
\be
{\rm Q}(r)=-2\, {\rm tr}\, ({\rm T}_3\, I)
=r^2\left(\frac{\eta(r)-2\chi(r)}{3r}\right)',
\ee
normalized such that Q$(\infty)=Q$. The behavior of Q$(r)$, 
shown in Fig.3,  suggests that part of the total charge $Q$
is hidden behind the horizon, and the rest is
distributed between the horizon and infinity.
If $r_h$ approaches zero, the hidden charge assumes a finite
value, such that its distribution becomes $\delta$-like, making
it plausible that there are no acceptable rotational 
perturbations for the solitons.

To summarize, we have found that there is a branch of 
slowly rotating colored black holes with an asymptotically
Abelian gauge field, whose electric flux is proportional to the 
angular momentum $J$. 
Mathematically  the appearance of the charge is understood
as follows: 
The static black holes, similarly to the Schwarzschild solution, 
admit in the far zone two independent hairs, describing 
rotational and charged excitations. However,  unlike the situation in the 
vacuum case, these two hairs become coupled to each other 
in the near zone via the background gauge field,
and regularity enforces a relation between $Q$ and $J$. 
Physically, on the other hand, this rotational 
``charging up" may seem rather
unusual, since one normally expects that rotation induces
only dipole corrections to the static field. --
Our second main result is that the particle-like static
Bartnik-McKinnon solutions do not admit 
continuously connected stationary rotational 
excitations. Despite this fact, we do not see any
reason why solitons in other matter models should not 
have rotational states. 
Rotating magnetic monopoles or Skyrmions might be
examples of such objects. 

\noindent
{\bf Acknowledgments.--} We would like to thank
Othmar Brodbeck for discussions, 
and also Marcus Heusler for discussions and the careful
reading of the manuscript. 
This work was supported by the Swiss National Science
Foundation.

\end{document}